\documentclass[11pt]{article}

\usepackage{latexsym}
\usepackage{amsmath}
\usepackage{graphicx}

\def\beq{\begin{eqnarray}}
\def\eeq{\end{eqnarray}}

\begin{document}

\fontsize{11}{14.5pt}\selectfont

\vspace*{20pt}

\begin{center} \LARGE \bf 
 Non-reversibly updating a uniform $[0,1]$ value \\ 
 for Metropolis accept/reject decisions\footnote{An earlier version of this work
 was presented at the 2nd Symposium on Advances in Approximate Bayesian
 Inference, Vancouver, British Columbia, 8 December 2019.}
\end{center}

\vspace{8pt}

\begin{center} 
 \large Radford M. Neal\footnote{
  Dept.\ of Statistical Sciences and Dept.\ of Computer Science,
  University of Toronto, Vector Institute Affiliate.
  \texttt{http://www.cs.utoronto.ca/$\sim$radford/},
  \texttt{radford@stat.utoronto.ca}}\\[4pt]
  31 January 2020
 
\end{center}

\vspace{12pt}

\noindent {\bf Abstract.}  I show how it can be beneficial to express
Metropolis accept/reject decisions in terms of comparison with a
uniform $[0,1]$ value, $u$, and to then update $u$ non-reversibly, as
part of the Markov chain state, rather than sampling it independently
each iteration.  This provides a small improvement for random walk
Metropolis and Langevin updates in high dimensions.  It produces a
larger improvement when using Langevin updates with persistent
momentum, giving performance comparable to that of Hamiltonian Monte
Carlo (HMC) with long trajectories.  This is of significance when some
variables are updated by other methods, since if HMC is used, these
updates can be done only between trajectories, whereas they can be
done more often with Langevin updates.  I demonstrate that for a
problem with some continuous variables, updated by HMC or Langevin
updates, and also discrete variables, updated by Gibbs sampling
between updates of the continuous variables, Langevin with persistent
momentum and non-reversible updates to $u$ samples nearly a factor of
two more efficiently than HMC.  Benefits are also seen for a Bayesian
neural network model in which hyperparameters are updated by Gibbs
sampling.

\section{Introduction}\vspace{-11pt}

In the Metropolis MCMC method, a decision to accept or reject a
proposal to move from $x$ to $x^*$ can be done by checking whether $u
< \pi(x^*)/\pi(x)$, where $\pi$ is the density function for the
distribution being sampled, and $u$ is a uniform $[0,1]$ random
variable.  Standard practice is to generate a value for $u$
independently for each decision.  I show here that it can be
beneficial to instead update $u$ each iteration without completely
forgetting the previous value, using a non-reversible method.

Doing non-reversible updates to $u$ will not change the fraction of
proposals that are accepted, but can result in acceptances being
clustered together (with rejections similarly clustered).  This can be
beneficial, for example, when rejections cause reversals of direction,
as in Horowitz's (1991) variant of Langevin updates for continuous
variables in which the momentum ``persists'' between updates.
Compared to the alternative of Hamiltonian Monte Carlo (HMC) with long
trajectories (Duane, Kennedy, Pendleton, and Roweth 1987; reviewed in
Neal 2011), Langevin updates allow other updates, such as for discrete
variables, to be done more often, which I will show can produce an
advantage over HMC if persistent momentum is used in conjunction with
non-reversible updates for $u$.


\section{The New Method for Accept/Reject Decisions}\vspace{-11pt}

For any MCMC method, we can augment the variable of interest, $x$,
with density $\pi(x)$, by a variable $s$, whose conditional
distribution given $x$ is uniform over $[0,\pi(x)]$.  The resulting
joint distribution for $x$ and $s$ is uniform over the region where $0
< s < \pi(x)$.  This is the view underlying ``slice sampling'' (Neal
2003), in which $s$ is introduced temporarily, by sampling uniformly
from $[0,\pi(x)]$, and then forgotten once a new $x$ has been chosen.
Metropolis updates can also be viewed in this way, with the new $x$
found by  accepting or rejecting a proposal, $x^*$, according to
whether $\pi(x^*) > s$, with $s$ newly sampled from $[0,\pi(x)]$.

However, it is valid to instead retain $s$ in the state between
updates, utilizing its current value for accept/reject decisions, and
updating this value when desired by any method that leaves invariant
the uniform distribution on $[0,\pi(x)]$, which is the conditional
distribution for $s$ given $x$ implied by the uniform distribution
over $(x,s)$ for which $0 < s < \pi(x)$.  

Equivalently, one can retain in the state a value $u$ whose
distribution is uniform over $[0,1]$, independent of $x$, with $u$
corresponding to $s/\pi(x)$ --- the state $(x,u)$ is just another way
of viewing the state $(x,s)$.  Accept/reject decisions are then made
by checking whether $u < \pi(x^*)/\pi(x)$.  Note, however, that if
$x^*$ is accepted, $u$ must then be updated to $u\, \pi(x) /
\pi(x^*)$, which corresponds to $s$ not changing.

Here, I will consider non-reversible updates for $u$, which translate
it by some fixed amount, $\delta$, and perhaps add some noise,
reflecting off the boundaries at $0$ and $1$.  It is convenient to
express such an update with reflection in terms of a variable $v$ that
is uniform over $[-1,+1]$, and define $u = |v|$.  An update for $v$
can then be done as follows:\vspace{-4pt}
\begin{eqnarray}
  \lefteqn{v\ \leftarrow\ v\ +\ \delta\ +\ \mbox{noise}} 
   & & \hspace*{1.5in} \nonumber \\
  \lefteqn{\mbox{while\ } v > +1:\ \ v\ \leftarrow\ v - 2} \label{eq-v-trans} \\
  \lefteqn{\mbox{while\ } v < -1:\ \ v\ \leftarrow\ v + 2}
   \nonumber \vspace{-4pt}
\end{eqnarray}
For any $\delta$ and any distribution for the noise (not depending on
the current value of $v$), this update leaves the uniform distribution
over $[-1,+1]$ invariant, since it is simply a wrapped translation,
which has Jacobian of magnitude one, and hence keeps the probability
density for the new $v$ the same as the old $v$ (both being $1/2$).

The full state consists of $x$ and $v$, with $x$ having density
$\pi(x)$ and, independently, $v$ being uniform over $[-1,+1]$, which
corresponds to the conditional distribution (given $x$) of $s=|v|\pi(x)$
being uniform over $[0,\pi(x)]$. If a proposed move from $x$ to $x^*$
is accepted we must change $v$ as follows:
\beq
  v & \leftarrow & v\,\pi(x)/\pi(x^*)
\label{eq-v-change}
\eeq
This leaves $s$ unchanged, and hence the reverse move would also
be accepted, as necessary for reversibility.  Because of this change to $v$
on acceptance, when $\pi(x)$ varies continuously, it may not be
necessary to include noise in the update for $v$ in (\ref{eq-v-trans}), 
but if $\pi(x)$ has only a finite number of possible values, adding noise 
may be necessary to ensure ergodicity.

The hope with these non-reversible updates is that $u$ will move
slowly (if $\delta$ and the noise amount are small) between values
near 0, where acceptance is easy, and values near 1, where acceptance
is harder.  (But note that $u$ may change in either direction when
proposals are accepted, though $s$ will not change.)  Non-reversibly
updating $u$ will not change the overall acceptance rate, since the
distribution of $u$ is still uniform over $[0,1]$, independent of $x$,
but it is expected that acceptances and rejections will become more
clustered --- with an accepted proposal more likely to be followed by
another acceptance, and a rejected proposal more likely to be followed
by another rejection.

We might wish to intermingle Metropolis updates for $x$ that use $v$
to decide on acceptance with other sorts of updates for $x$ --- for
example, Gibbs sampling updates, or Metropolis updates accepted in the
standard fashion.  We can do these updates while ignoring $v$, and
then simply resume use of $v$ afterwards, since $v$ is independent of
$x$.  We could indeed include several independent $v$ variables in the
state, using different $v$ values for different classes of updates,
but this generalization is not explored here.

\section{Results for random-walk Metropolis updates in high 
         dimensions}\vspace{-11pt}\label{sec-res-met-high-dim}

A small benefit from non-reversible updating of $u$ can be seen with
simple random-walk Metropolis updates.  Such updates operate as 
follows:\vspace{-4pt}
\begin{enumerate}
\item[1)] Propose $x^* \sim N(x,\sigma^2I)$, where $\sigma$ is a
      stepsize parameter that needs to be tuned.\vspace{-4pt}
\item[2)] Accept $x^\prime = x^*$ as next state if
      $u < \pi(x^*)/\pi(x)$,; otherwise let $x^\prime=x$.\vspace{-4pt}
\end{enumerate}

\begin{figure}[p]

\vspace*{-20pt}
\hspace*{20pt}\includegraphics[scale=1,angle=-90]{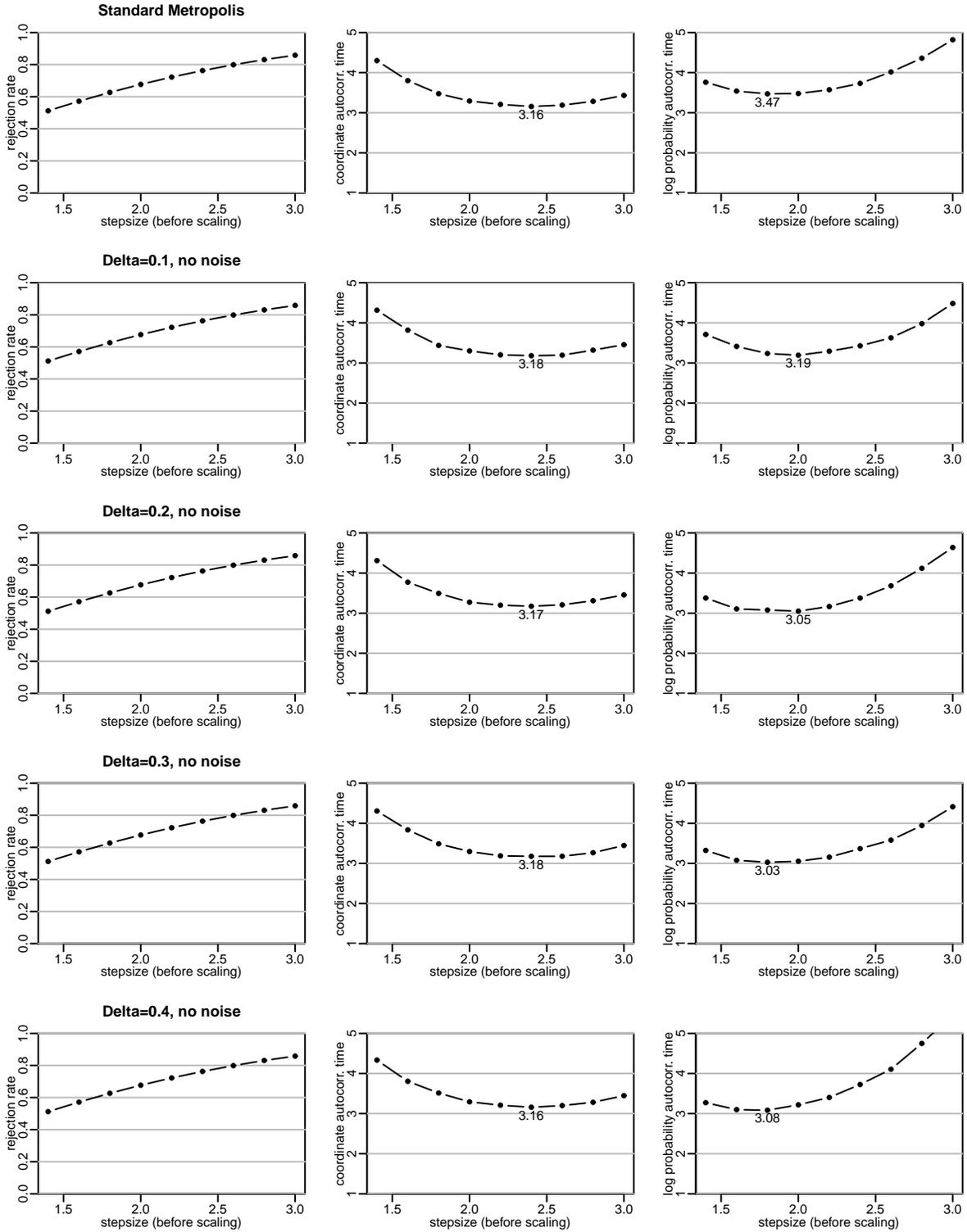}

\caption{Sampling from a 40-dimensional Gaussian with identity covariance.
The values for $\sigma$ used were the stepsizes shown above divided by
$40^{1/2}$.  The autocorrelation times (one plus twice the sum
of autocorrelations at lags 1 to when they are negligible)
that are shown are for groups of 40 iterations (hence the
single-iteration autocorrelation time is about 40 times larger).
Results are shown for standard Metropolis, and for 
Metropolis with non-reversible update for $u$, with $\delta=0.1, 0.2, 0.3, 0.4$,
and no noise.}\label{fig-met}

\end{figure}

Figure~\ref{fig-met} compares results of such updates when $u$ is
sampled independently each iteration versus being updated
non-reversibly as described in Section 2, when sampling a
40-dimensional Gaussian distribution with identity covariance matrix.
When estimating the mean of a single coordinate, little difference is
seen, but for estimating the mean of the log of the probability
density, the non-reversible method with $\delta=0.3$ is $3.47/3.03
\approx 1.15$ times better.

One possible explanation for the better performance with 
non-reversible updates is that, as noted by Caracciolo, et al (1994),
the performance of Metropolis methods in high dimensions is limited by
their ability to sample different values for the log of the density.
In $D$ dimensions, the log density typically varies over a range
proportional to $D^{1/2}$.  A Metropolis update will typically change
the log density only by about one --- larger decreases in the log
density are unlikely to be accepted, and it follows from reversibility
that increases in the log density of much more than one must also be
rare (once equilibrium has been reached).  Since standard Metropolis
methods are reversible, these changes of order one will occur in a
random walk, and so around $D$ steps will be needed to traverse the
range of log densities of about $D^{1/2}$, limiting the speed of
mixing.

The gain seen from using non-reversible updates for $u$ may come from
helping with this problem.  When $u$ is small few proposals will be
rejected, and the chain will tend to drift towards smaller values for
the log density, with the opposite behaviour at times when $u$ is near
one.  This could reduce the random walk nature of changes in the log
density.  Note, however, that this effect is limited by the changes
to $v$ (and hence $u$) required by equation~(\ref{eq-v-change}).

\section{Application to Langevin updates with persistent momentum}\vspace{-11pt}

I obtained more interesting results when using non-reversible updates
of $u$ in conjunction with the one-step, non-reversible version of
Hamiltonian Monte Carlo (Duane, et al 1987) due to Horowitz (1991).
This method is a form of ``Langevin'' update in which the momentum is
only partially replaced in each iteration, so that the method tends to
move persistently in the same direction for many iterations.  I
briefly present this method below; it is discussed in more detail in
my review of Hamiltonian Monte Carlo methods (Neal 2011, Section 5.3).

Hamiltonian Monte Carlo (HMC) works in a state space extended with momentum
variables, $p$, of the same dimension as $x$, which are newly sampled each 
iteration.  The distribution for $x$
is defined in terms of a ``potential energy'' function $U(x)$, with
$\pi(x) \propto \exp(-U(x))$.  HMC proposes a new value 
for $(x,p)$ by simulating Hamiltonian dynamics with some number of
``leapfrog'' steps (and then negating $p$, so the proposal is
reversible).  A leapfrog step from $(x_t,p_t)$ to $(x_{t+\eta},p_{t+\eta})$
has the form\vspace{-4pt}
\begin{eqnarray}
 p_{t+\eta/2} & = & p_t\ -\ (\eta/2)\,\nabla U(x_t) \nonumber \\
 x_{t+\eta}   & = & x_t\ +\ \eta\,p_{t+\eta/2} \\
 p_{t+\eta}   & = & p_{t+\eta/2}\ -\ (\eta/2)\,\nabla U(x_{t+\eta})
  \nonumber \vspace{-4pt}
\end{eqnarray}
$L$ such steps may be done to go from the current state
$(x,p) = (x_0,p_0)$ to the proposal
$(x^*,p^*)=(x_{\eta L},-p_{\eta L})$.  If $\eta L$ is large, but $\eta$ 
is sufficiently small (so acceptance is likely),
a proposed point can be accepted that is distant from the current point,
avoiding the slowdown from doing a random walk with small steps.  In
standard HMC, $p$ is sampled independently at each iteration from 
the $N(0,I)$ distribution, randomizing the direction in which the next 
trajectory starts out.

The standard Langevin method is equivalent to HMC with $L=1$.  This
method makes use of gradient information, $\nabla U(x)$, which gives
it an advantage over simple random-walk Metropolis methods, but it
does not suppress random walks, since $p$ is randomized after each 
step.

In Horowitz's method, only one leapfrog step is done each iteration,
as for standard Langevin updates, but a trick is used so that these
steps nevertheless usually keep going in the same direction for many
iterations, except on a rejection.  The updates with this method
operate as follows:\vspace{-4pt}

\begin{enumerate}
\item[1)]
      Set $p^\prime = \alpha p\ +\, \sqrt{1\!-\!\alpha^2}\, n$,
      where $n \sim N(0,I)$, and $\alpha$ is a tuning parameter.
      Set $x^\prime=x$.\vspace{-4pt}
\item[2)] Find $(x^*\!,p^*)$ from $(x^\prime\!,p^\prime)$ with one
      leapfrog step, with stepsize $\eta$, followed
      by negation of $p$.\vspace{-4pt}
\item[3)] Accept $(x^{\prime\prime},\!p^{\prime\prime}) = (x^*\!,p^*)$
   if $u\ <\ \pi(x^*\!,p^*)/\pi(x^\prime\!,p^\prime)$;
   otherwise
   $(x^{\prime\prime}\!,p^{\prime\prime})=(x^\prime\!,p^\prime)$.\vspace{-4pt}
\item[4)]Let $p^{\prime\prime\prime} = -p^{\prime\prime}$ and
      $x^{\prime\prime\prime}=x^{\prime\prime}$.\vspace{-4pt}
\end{enumerate}

For $\alpha$ slightly less than 1, Step (1) only slightly changes $p$.
If Step (3) accepts, the negation in the proposal is canceled by the
negation in Step (4).  But a rejection will reverse $p$, leading the
chain to almost double back on itself.  Setting $\alpha$ to zero
gives the equivalent of the standard Langevin method.

Unfortunately, even with this non-reversibility trick, and an
optimally-chosen $\alpha$, Horowitz's method is not as efficient as
HMC with long trajectories.  To reduce the rejection rate, and hence
random-walk-inducing reversals of direction, a small, inefficient
stepsize ($\eta$) is needed.

But a higher rejection rate would be tolerable if rejections cluster
in time, producing long runs of no rejections.  For example,
rejection-free runs of 20, 0, 20, 0, \ldots are better than
rejection-free runs of 10, 10, 10, 10, \ldots, since $N$ of the former
runs will potentially move via a random walk a distance proportional
to $20(N/2)^{1/2}\, \approx\, 14\,N^{1/2}$, whereas $N$ of the latter
runs will move only a distance proportional to $10\,N^{1/2}$.

We may hope that such clustering of rejections will be obtainable
using non-reversible updates of $u$, with consequent improvement in
the performance of Langevin with persistent momentum.\footnote{Non-reversible
updates of $u$ can also benefit the standard Langevin method. In
particular, when sampling from the 40-dimensional Gaussian of
Section~\ref{sec-res-met-high-dim}, using non-reversible updates of $u$
(with $\delta=0.7$) results in a factor of 1.19 lower autocorrelation time
for the log probability density compared to the standard Langevin method.}

\section{Results for Langevin updates with persistent 
         momentum}\label{sec-res-plang}\vspace{-11pt}

To test whether non-reversible updating of $u$ can improve the
Langevin method with persistent momentum, I first tried sampling a
multivariate Gaussian distribution consisting of 16 independent pairs
of variables having variances of 1 and correlations of 0.99 (i.e., a
32-dimensional Gaussian with block-diagonal covariance matrix).  For
this test, the methods are not set up to use knowledge of this
correlation structure, mimicking problems in which the dependencies
take a more complex, and unknown, form.

The high correlation within each pair and (moderately) high
dimensionality limit the stepsize ($\eta$) that can be used (while
avoiding a high rejection rate).  This test is therefore
representative of problems in which sampling is much more efficient
when many consecutive small steps to go in the same direction, rather
than doing a random walk.

\begin{figure}[b]

\vspace*{-17pt}
\hspace*{-5pt}\includegraphics[scale=0.75,
                                angle=-90]{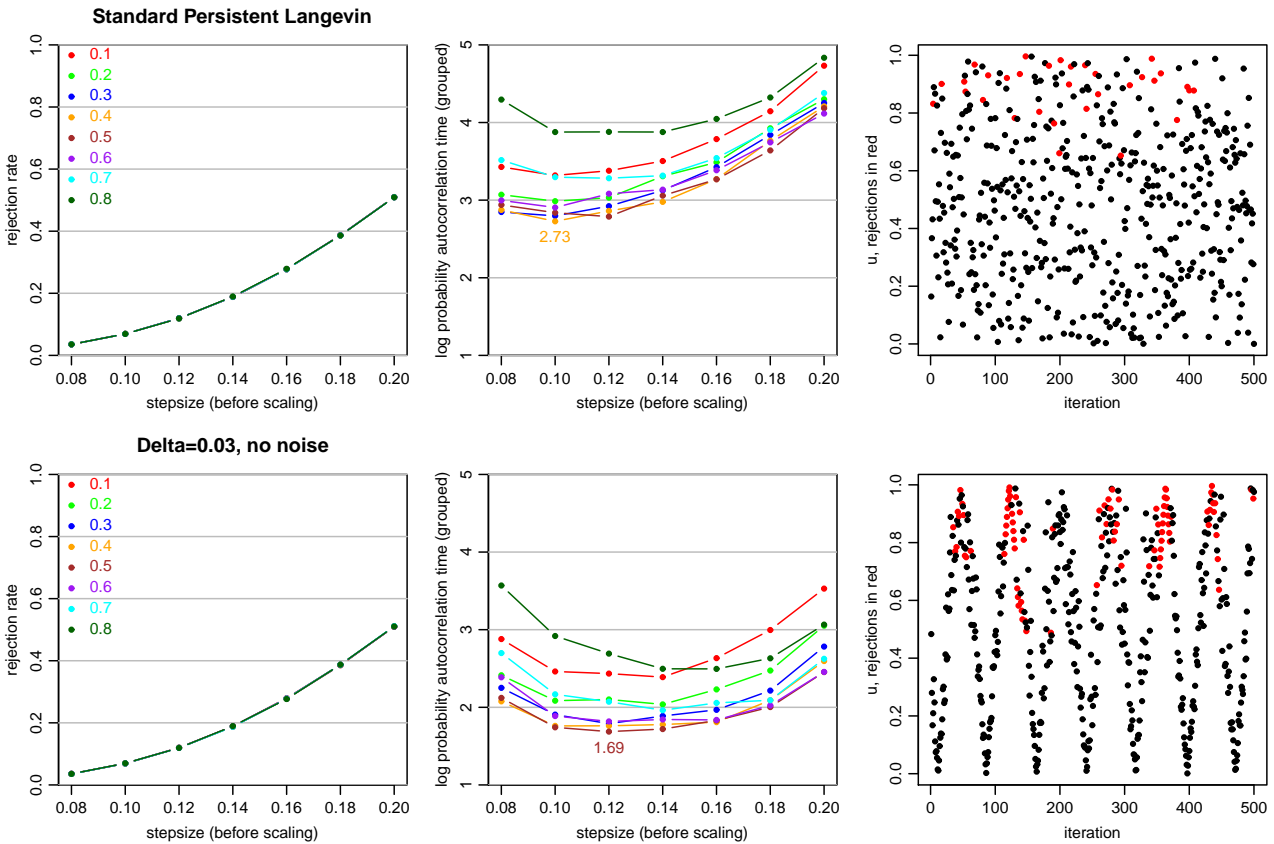}

\caption{Sampling from a 32-dimensional Gaussian distribution consisting
of independent pairs of correlated variables with persistent Langevin
methods.  The values of $\eta$ used
were the stepsizes shown divided by $32^{1/6}$.  The values of $\alpha$
used were the colour-coded values on the left raised to the power $\eta$.
The autocorrelation times shown are for groups of 31 iterations.
The plots on the right show the values of $u$ used for 500 individual Langevin
updates, with the parameters that give the lowest autocorrelation time
for the log probability density, with values that led to rejection shown in
red.\vspace{-3pt}}\label{fig-plang}

\end{figure}

Figure~\ref{fig-plang} shows the results.  The centre plots show that
the run using a non-reversible update for $u$ (with $\delta=0.03$,
$\eta=0.12/32^{1/6}=0.0673$, and $\alpha=0.5^{\eta}=0.954$) is
$2.73/1.69 \approx 1.62$ times better at estimating the mean of the
log probability density than the best run of the standard persistent
Langevin method (with $\eta=0.10/32^{1/6}=0.0561$ and
$\alpha=0.4^{\eta}=0.950$).  The right plots show that rejections are
indeed clustered when non-reversible updates are used, which reduces
random walks.

For comparison, I tried HMC for this problem, with stepsizes of
$\eta=0.04,0.05,\ldots,0.14,0.15$ and trajectories with the number of
leapfrog steps being $L=1,2,4,8,16,32$.  The autocorrelation time for
the log probability density was found for groups of $32/L$ HMC
updates, giving results that are roughly comparable to those shown in
Figure 2.  The best performance was for $L=16$ and $\eta=0.07$, for
which the autocorrelation time was 2.04 (though it's possible that
better results might be found with a search over a finer grid for
$L$).  This is worse than the best result of 1.69 for persistent
Langevin with non-reversible updates of $u$, using the stepsize of
$0.12/32^{1/6} = 0.0673$, which interestingly is close to the optimal
stepsize for HMC.  Use of a non-reversible update for $u$ has allowed
persistent Langevin to match or exceed the performance of HMC for this
problem.

I also used these methods to sample from a mixed continuous/discrete
distribution, which may be specified as follows (with variables 
being conditionally independent except as shown):\vspace{-3pt}
\begin{eqnarray*}
  u & \sim & N(0,1) \\
  v\ |\ u & \sim & N(u,\,0.04^2) \\
  w_1\ |\ u & \sim & \mbox{Bernoulli\,}(1/(1+e^u)) \\
  & \vdots & \\
  w_{20}\ |\ u & \sim & \mbox{Bernoulli\,}(1/(1+e^u))
\end{eqnarray*}
This distribution is intended to mimic more complex distributions
having both continuous and discrete variables, which are dependent.
I will consider only MCMC methods that alternate updates for the
continuous variables, $u$ and $v$, with updates for the binary variables,
$w_1,\ldots,w_{20}$ (by Gibbs sampling).

Since the marginal distribution of $u$ is known to be $N(0,1)$, the
quality of the sampling for $u$ is easily checked.  Because of the
high dependence between $u$ and the 20 binary variables, we might
expect that a method that frequently updates both the continuous and
the binary variables will do better than a method, such as HMC with long
trajectories, in which the binary variables are updated infrequently
--- i.e., only after substantial computation time is spent updating
the continuous variables.  

In assessing performance, I assume that computation time is dominated
by updates of the continuous variables, which in a real
application would be more numerous, and in particular by evaluation of
their log density, conditional on the current values of the binary
variables.  (This is not always actually the case here, however, due
to the problem simplicity and to avoidable inefficiencies in the
software used.)

I tried HMC with trajectories of 30, 40, or 60 leapfrog steps ($L$),
and stepsize ($\eta$) of 0.025, 0.030, 0.035, 0.040, and 0.045.  Gibbs
sampling updates were done between trajectories.
Autocorrelation times were found for a group of $120/L$ HMC
trajectories.  Persistent Langevin with non-reversible updates of $u$
was tried with $\delta$ of 0.003, 0.005, 0.010, and 0.015, each with
all combinations of $\alpha$ in 0.98, 0.99, 0.995, 0.9975, 0.9985, and
0.9990 and $\eta$ in 0.015, 0.020, 0.025, 0.030, 0.040, and 0.050.
Gibbs sampling updates were done after every tenth update.
Autocorrelation times were found for a group of 60 Langevin updates
--- note that this involves half as many log density gradient
evaluations as for a group of HMC iterations.

Performance was evaluated in terms of estimating the mean of the
indicator for $u$ being in $(-0.5,1.5)$.  Autocorrelation time for
this function was minimized for HMC with $L=40$ and $\eta=0.035$, with
the minimum being 1.53.  Autocorrelation time was minimized for
persistent Langevin with non-reversible updates of $u$ with
$\delta=0.010$, $\alpha=0.995$, and $\eta=0.030$, with the minimum
being 1.67.  Considering that the HMC groups require twice as many
gradient evaluations, persistent Langevin with non-reversible updates
of $u$ is $2 \times 1.53\,/\,1.67 = 1.83$ times more efficient than HMC
for this problem.

\section{Use for Bayesian neural networks}\vspace{-11pt}

I have also tried using the persistent Langevin method with
non-reversible updates for $u$ to sample the posterior distribution of
a Bayesian neural network model.  Such models (Neal 1995) typically
have hyperparameters controlling the variance of groups of weights in
the network.  It is convenient to use Gibbs sampling updates for these
hyperparameters, alternating such updates with HMC updates for the
network weights.  However, when long trajectories are used for HMC, as
is desirable to reduce random-walk behaviour, the Gibbs sampling
updates for hyperparameters are done infrequently.  Using persistent
Langevin updates for weights would allow hyperparameters to be updated
more frequently, perhaps speeding overall convergence.  We may hope
that this will work better with a non-reversible update for $u$.

I tested this approach on a binary classification problem, with 5
inputs, and 300 training cases.  A network with one hidden layer of 12
tanh hidden units was used.  Only two of the inputs for this problem
were relevant, with two more being slightly noisy versions of the two
relevant inputs, and one input being independent noise.  Five separate
hyperparameters controlled the variances of weights out of each input.

For the best-tuned persistent Langevin method with non-reversibe
update of $u$, the average autocorrelation time for the four
plausibly-relevant input hyperparameters was 1.25 times smaller than
for the best-tuned HMC method.  This is an encouraging preliminary
result, but tests with more complex networks are needed to assess how
much Bayesian neural network training can be improved in this way.

\section{References}\vspace{-11pt}

\leftmargini 0.2in

\begin{description}

\item[\hspace{-5pt}]
  Caracciolo, S, Pelisseto, A, and Sokal, A.~D.\ (1994) ``A general limitation
  on Monte Carlo algorithms of Metropolis type'', {\em Physical Review Letters},
  vol.~72, pp.~179-182.  Also available 
  at \texttt{arxiv.org/abs/hep-lat/9307021}

\item[\hspace{-5pt}]
  Duane, S., Kennedy, A.~D., Pendleton, B.~J., and Roweth, D.\ (1987)
  ``Hybrid Monte Carlo'', {\em Physics Letters B}, vol.~195, pp.~216-222.


\item[\hspace{-5pt}]
  Horowitz, A.~M.\ (1991) ``A generalized guided Monte Carlo algorithm'',
  \textit{Physics Letters B}, vol.~268, pp.~247-252.

\item[\hspace{-5pt}]
  Neal, R.~M.\ (2003) ``Slice sampling'' (with discussion), {\em Annals of 
  Statistics}, vol.~31, pp.~705-767.

\item[\hspace{-5pt}]
  Neal, R.~M.\ (2011) ``MCMC using Hamiltonian dynamics'', in the
  \textit{Handbook of Markov Chain Monte Carlo}, S.\ Brooks, A.\ Gelman,
  G.\ L.\ Jones, and X.-L.\ Meng (editors), Chapman \& Hall / CRC Press,
  pp.~113-162. Also available at \texttt{arxiv.org/abs/1206.1901}

\item[\hspace{-5pt}]
  Neal, R.~M.\ (1995) {\em Bayesian Learning for Neural Networks},
  Ph.D.\ Thesis, Dept.\ of Computer Science, University of Toronto, 195 pages.

\end{description}

\pagebreak

\section*{Appendix --- Scripts used to produce results}\vspace{-11pt}

The experimental results reported were obtained using my Software for
Flexible Bayesian Modeling (FBM, the release of 2020-01-24), available
at \texttt{www.cs.utoronto.ca/$\sim$radford/fbm.software.html} or at
\texttt{gitlab.com/radfordneal/fbm}.  This software takes the form of
a number of programs that are run using Unix/Linux bash shell scripts.
Here, I will show scripts that can be used to produce single points in
the figures, omitting the more complex scripts and programs used to
run the methods with multiple parameter settings and plot the results.

\subsection*{Scripts for random-walk Metropolis in high dimensions}\vspace{-8pt}

For Figure~\ref{fig-met}, the following commands may be used to produce 
the results for standard Metropolis sampling with stepsize of 1.8
(before scaling):

{\small \begin{verbatim}
        dim=40
        stepunscaled=1.8
        step=`calc "$stepunscaled/Sqrt($dim)"`
        len=1001000
        
        echo "STANDARD METROPOLIS"
        
        log=fig-met.log
        bvg-spec $log 1 1 0 `calc $dim/2`
        mc-spec $log repeat $dim metropolis $step
        bvg-mc $log $len
        
        echo -n "Rejection Rate: "
        bvg-tbl r $log 1001: | series m | tail -1 | sed "s/.*mean://"
        echo -n "Autocorrelation time for coordinate: "
        bvg-tbl q1 $log 1001: | series mvsac 10 0 | tail -1 | sed "s/.* //"
        echo -n "Autocorrelation time for log prob: "
        bvg-tbl E $log 1001: | series mvsac 10 `calc $dim/2` | tail -1 | sed "s/.* //"
\end{verbatim}}
\noindent which produce the following output:

{\small \begin{verbatim}
        STANDARD METROPOLIS
        Rejection Rate:  0.626588  
        Autocorrelation time for coordinate: 3.475440
        Autocorrelation time for log prob: 3.470835
\end{verbatim}}

The results in Figure~\ref{fig-met} for Metropolis updates with
non-reversible updates of $u$, with stepsize of 1.8 (before scaling)
and $\delta$ of 0.3, may be produced with the following commands:

{\small \begin{verbatim}
        dim=40
        stepunscaled=1.8
        step=`calc "$stepunscaled/Sqrt($dim)"`
        delta=0.3
        len=1001000
        
        echo "METROPOLIS WITH NON-REVERSIBLE UPDATE OF U"
        
        log=fig-met-nrevu.log
        bvg-spec $log 1 1 0 `calc $dim/2`
        mc-spec $log slevel $delta repeat $dim metropolis $step
        bvg-mc $log $len
        
        echo -n "Rejection Rate: "
        bvg-tbl r $log 1001: | series m | tail -1 | sed "s/.*mean://"
        echo -n "Autocorrelation time for coordinate: "
        bvg-tbl q1 $log 1001: | series mvsac 10 0 | tail -1 | sed "s/.* //"
        echo -n "Autocorrelation time for log prob: "
        bvg-tbl E $log 1001: | series mvsac 10 `calc $dim/2` | tail -1 | sed "s/.* //"
\end{verbatim}}
\noindent which produce the following output:

{\small \begin{verbatim}
        METROPOLIS WITH NON-REVERSIBLE UPDATE OF U
        Rejection Rate:  0.626545  
        Autocorrelation time for coordinate: 3.487568
        Autocorrelation time for log prob: 3.028137
\end{verbatim}}

In both scripts, 1001000 groups of 40 updates are simulated, with the
first 1000 being discarded as burn-in, and the remainder used to
estimate rejection rate and autocorrelation times.  The
autocorrelation times are found using estimated autocorrelations up to
lag 10, with the true means for a coordinate and for the log
probability density being used for these estimates.  (Note: Rather
than the log probability density itself, the ``energy'' is used, which
gives equivalent results since it is a linear function of the log
probability density.)

\subsection*{Scripts for persistent Langevin and HMC for replicated
bivariate Gaussian}\vspace{-8pt}

For Figure~\ref{fig-plang}, the following commands may be used to produce 
the results for standard persistent Langevin sampling with stepsize of 0.10
(before scaling) and an $\alpha$ value derived from 0.4:

{\small \begin{verbatim}
        dim=32
        stepunscaled=0.10
        alphabase=0.4
        step=`calc "$stepunscaled/Exp(Log($dim)/6)"`
        rep=`calc "v=10*Exp(Log($dim)/3)" "v-Frac(v)"`
        decay=`calc "Exp(Log($alphabase)*$step)"`
        len=101000
        
        echo "STANDARD PERSISTENT LANGEVIN"
        
        log=fig-plang.log
        bvg-spec $log 1 1 0.99 `calc $dim/2`
        mc-spec $log repeat $rep heatbath $decay hybrid 1 $step negate
        
        bvg-mc $log $len
        echo -n "Rejection Rate "
        bvg-tbl r $log 1001: | series m | tail -1 | sed "s/.*mean://"
        echo -n "Autocorrelation time for coordinate: "
        bvg-tbl q1 $log 1001: | series vsac 10 0 | tail -1 | sed "s/.* //"
        echo -n "Autocorrelation time for log prob: "
        bvg-tbl E $log 1001: | series vsac 10 `calc $dim/2` | tail -1 | sed "s/.* //"
\end{verbatim}}
\noindent which produce the following output:

{\small \begin{verbatim}
        STANDARD PERSISTENT LANGEVIN
        Rejection Rate  0.069295  
        Autocorrelation time for coordinate: 6.875574
        Autocorrelation time for log prob: 2.727262
\end{verbatim}}

The results in Figure~\ref{fig-plang} for persistent Langevin updates
with non-reversible updates of $u$, with stepsize of 0.12 (before
scaling), an $\alpha$ value derived from 0.5, and $\delta$ of 0.03,
may be produced with the following commands:

{\small \begin{verbatim}
        dim=32
        stepunscaled=0.12
        alphabase=0.5
        step=`calc "$stepunscaled/Exp(Log($dim)/6)"`
        rep=`calc "v=10*Exp(Log($dim)/3)" "v-Frac(v)"`
        decay=`calc "Exp(Log($alphabase)*$step)"`
        delta=0.03
        len=101000
        
        echo "PERSISTENT LANGEVIN WITH NON-REVERSIBLE UPDATE OF U"
        
        log=fig-plang.log
        bvg-spec $log 1 1 0.99 `calc $dim/2`
        mc-spec $log slevel $delta repeat $rep heatbath $decay hybrid 1 $step negate
        
        bvg-mc $log $len
        echo -n "Rejection Rate "
        bvg-tbl r $log 1001: | series m | tail -1 | sed "s/.*mean://"
        echo -n "Autocorrelation time for coordinate: "
        bvg-tbl q1 $log 1001: | series vsac 10 0 | tail -1 | sed "s/.* //"
        echo -n "Autocorrelation time for log prob: "
        bvg-tbl E $log 1001: | series vsac 10 `calc $dim/2` | tail -1 | sed "s/.* //"
\end{verbatim}}
\noindent which produce the following output:

{\small \begin{verbatim}
        PERSISTENT LANGEVIN WITH NON-REVERSIBLE UPDATE OF U
        Rejection Rate  0.119244  
        Autocorrelation time for coordinate: 2.827302
        Autocorrelation time for log prob: 1.686796
\end{verbatim}}

Note that autocorrelation times for coordinates are not plotted in
Figure~\ref{fig-plang} because, due to the ability of persistent
Langevin methods to produce negative correlations, the coordinate
autocorrelation times approach zero as $\eta$ goes to zero and
$\alpha$ goes to one.  So there is no optimal choice of parameters by
this criterion (which in this limit does not represent practical
utility).

The HMC comparison results were obtained with commands such as the
following (for stepsize of 0.07 and 16 leapfrog steps):

{\small \begin{verbatim}
        dim=32
        step=0.07
        leaps=16
        rep=`calc 32/$leaps`
        len=101000
        
        echo "HMC"
        
        log=hmc-compare.log
        
        bvg-spec $log 1 1 0.99 `calc $dim/2`
        mc-spec $log repeat $rep heatbath hybrid $leaps $step:30
        
        bvg-mc $log $len
        echo -n "Rejection Rate "
        bvg-tbl r $log 1001: | series m | tail -1 | sed "s/.*mean://"
        echo -n "Autocorrelation time for coordinate: "
        bvg-tbl q1 $log 1001: | series vsac 10 0 | tail -1 | sed "s/.* //"
        echo -n "Autocorrelation time for log prob: "
        bvg-tbl E $log 1001: | series vsac 10 `calc $dim/2` | tail -1 | sed "s/.* //"
\end{verbatim}}
\noindent which produce the following output:
{\small \begin{verbatim}
        HMC
        Rejection Rate  0.142875  
        Autocorrelation time for coordinate: 3.364492
        Autocorrelation time for log prob: 2.038866
\end{verbatim}}

In this script, in order to avoid possible periodicity effects, the
leapfrog stepsize for HMC is randomized for each trajectory by
multiplying the nominal stepsize by the reciprocal of the square root
of a Gamma random variable with mean 1 and shape parameter 30/2.

\subsection*{Scripts for persistent Langevin and HMC for mixed 
             continuous/discrete distribution}\vspace{-8pt}

The results in Section~\ref{sec-res-plang} for persistent Langevin with
non-reversible update of $u$ applied to a mixed continuous/discrete
distribution, with $\delta=0.010$, $\alpha=0.995$, and $\eta=0.030$,
may be produced with the following commands:

{\small \begin{verbatim}
        delta=0.010
        alpha=0.995
        step=0.030
        
        echo "PERSISTENT LANGEVIN / NON-REVERSIBLE UPDATE OF U FOR MIXED DISTRIBUTION"
        
        log=mixed-plang.log
        
        dist-spec $log "u ~ Gaussian(0,1) + v ~ Gaussian(u,0.04^2) + \
                        y0 ~ Bernoulli(1/(1+Exp(u))) + \
                        y1 ~ Bernoulli(1/(1+Exp(u))) + \
                        y2 ~ Bernoulli(1/(1+Exp(u))) + \
                        y3 ~ Bernoulli(1/(1+Exp(u))) + \
                        y4 ~ Bernoulli(1/(1+Exp(u))) + \
                        y5 ~ Bernoulli(1/(1+Exp(u))) + \
                        y6 ~ Bernoulli(1/(1+Exp(u))) + \
                        y7 ~ Bernoulli(1/(1+Exp(u))) + \
                        y8 ~ Bernoulli(1/(1+Exp(u))) + \
                        y9 ~ Bernoulli(1/(1+Exp(u))) + \
                        z0 ~ Bernoulli(1/(1+Exp(u))) + \
                        z1 ~ Bernoulli(1/(1+Exp(u))) + \
                        z2 ~ Bernoulli(1/(1+Exp(u))) + \
                        z3 ~ Bernoulli(1/(1+Exp(u))) + \
                        z4 ~ Bernoulli(1/(1+Exp(u))) + \
                        z5 ~ Bernoulli(1/(1+Exp(u))) + \
                        z6 ~ Bernoulli(1/(1+Exp(u))) + \
                        z7 ~ Bernoulli(1/(1+Exp(u))) + \
                        z8 ~ Bernoulli(1/(1+Exp(u))) + \
                        z9 ~ Bernoulli(1/(1+Exp(u)))"
        
        mc-spec $log slevel $delta \
                 repeat 6 repeat 10 heatbath $alpha hybrid 1 $step 0:1 negate end \
                        binary-gibbs 2:21
        dist-mc $log 200000
        
        echo -n "Rejection Rate: "
        dist-tbl r $log 1001: | series m | tail -1 | sed "s/.*mean://"
        
        echo -n "Autocorrelation time for I(-0.5,1.5): "
        (echo "u<-scan(quiet=T)"; \
         dist-tbl u $log 1001:; echo ""; \
         echo "write.table(as.numeric(u>-0.5&u<1.5),row.names=F,col.names=F)" \
        ) | R --slave --vanilla | series vsac 15 0.6246553 | tail -1 | sed "s/.* //"
\end{verbatim}}
\noindent which produce the following output:
{\small \begin{verbatim}
        PERSISTENT LANGEVIN / NON-REVERSIBLE UPDATE OF U FOR MIXED DISTRIBUTION
        Rejection Rate:  0.093834  
        Autocorrelation time for I(-0.5,1.5): 1.666017
\end{verbatim}}
\noindent
Note that each group of iterations performs 60 leapfrog steps.

The results for HMC, with 40 leapfrog steps and $\eta=0.035$,
may be produced with the following commands:
{\small \begin{verbatim}
        leaps=40
        step=0.035
        rep=`calc 120/$leaps`
        
        echo "HMC FOR MIXED DISTRIBUTION"
        
        log=mixed-hmc.log
        
        dist-spec $log "u ~ Gaussian(0,1) + v ~ Gaussian(u,0.04^2) + \
                        y0 ~ Bernoulli(1/(1+Exp(u))) + \
                        y1 ~ Bernoulli(1/(1+Exp(u))) + \
                        y2 ~ Bernoulli(1/(1+Exp(u))) + \
                        y3 ~ Bernoulli(1/(1+Exp(u))) + \
                        y4 ~ Bernoulli(1/(1+Exp(u))) + \
                        y5 ~ Bernoulli(1/(1+Exp(u))) + \
                        y6 ~ Bernoulli(1/(1+Exp(u))) + \
                        y7 ~ Bernoulli(1/(1+Exp(u))) + \
                        y8 ~ Bernoulli(1/(1+Exp(u))) + \
                        y9 ~ Bernoulli(1/(1+Exp(u))) + \
                        z0 ~ Bernoulli(1/(1+Exp(u))) + \
                        z1 ~ Bernoulli(1/(1+Exp(u))) + \
                        z2 ~ Bernoulli(1/(1+Exp(u))) + \
                        z3 ~ Bernoulli(1/(1+Exp(u))) + \
                        z4 ~ Bernoulli(1/(1+Exp(u))) + \
                        z5 ~ Bernoulli(1/(1+Exp(u))) + \
                        z6 ~ Bernoulli(1/(1+Exp(u))) + \
                        z7 ~ Bernoulli(1/(1+Exp(u))) + \
                        z8 ~ Bernoulli(1/(1+Exp(u))) + \
                        z9 ~ Bernoulli(1/(1+Exp(u)))"
        
        mc-spec $log repeat $rep heatbath hybrid $leaps $step:10 0:1 binary-gibbs 2:21
        dist-mc $log 200000
        
        echo -n "Rejection Rate: "
        dist-tbl r $log 1001: | series m | tail -1 | sed "s/.*mean://"
        
        echo -n "Autocorrelation time for I(-0.5,1.5): "
        (echo "u<-scan(quiet=T)"; \
         dist-tbl u $log 1001:; echo ""; \
         echo "write.table(as.numeric(u>-0.5&u<1.5),row.names=F,col.names=F)" \
        ) | R --slave --vanilla | series vsac 15 0.6246553 | tail -1 | sed "s/.* //"
\end{verbatim}}
\noindent which produce the following output:
{\small \begin{verbatim}
        HMC FOR MIXED DISTRIBUTION
        Rejection Rate:  0.171698  
        Autocorrelation time for I(-0.5,1.5): 1.527655
\end{verbatim}}
\noindent
Note that each group of iterations performs 120 leapfrog steps, and hence
(is assumed to) take twice as long as for the persistent Langevin script.

\end{document}